# Machine Learning Assisted NEO Discovery and Polarimetric Characterisation with Astronomical Surveys

G.A. Verdoes Kleijn[1], T. Grobler[2], S.J. Chong[1,2], O.R. Williams[1], M. Micheli[3], D. Koschny[4,5], T. Saifollahi[6], L.V.E. Koopmans[1], D. Dirkx[7], T. Santana-Ros[8,9], Y.-Z. Ma[2], M. Pöntinen[10], S. Bagnulo[11], M. Granvik[10,12], B.Y. Irureta-Goyena[13]

[1] *University of Groningen, postbus 800, 9700 AV, Groningen, The Netherlands*
[2] *Stellenbosch University, Private Bag X1, 7602 Matieland, South Africa*
[3] *ESA Near-Earth Object Coordination Centre, ESRIN, Largo Galileo Galilei 1 00044, Frascati (Rome), Italy*
[4] *Lunar and Planetary Exploration, LPE/TUM, Lise-Meitner-Str. 9 85521, Ottobrun, Germany*
[5] *Space Exploration Institute (SPACE-X), Fbg de l'Hôpital 68, 2000 Neuchâtel, Switzerland*
[6] *Observatoire Astronomique de Strasbourg, 11 Rue de l'Université, 67000 Strasbourg, France.*
[7] *Delft University of Technology, Mekelweg 5, 2628 CD, Delft, The Netherlands.*
[8] *Universidad de Alicante, Carr. de San Vicente del Raspeig, s/n, 03690 San Vicente del Raspeig, Alicante, Spain*
[9] *Universitat de Barcelona, Carrer de Martí i Franquès, 1, 08028 Barcelona, Spain*
[10] *Department of Physics, PO Box 64, 00014 University of Helsinki, Finland*
[11] *Armagh Observatory and Planetarium: Armagh, GB*
[12] *Division of Space Technology, Luleå University of Technology, Box 848, 98128 Kiruna, Sweden*
[13] *Laboratory of Astrophysics, École Polytechnique Fédérale de Lausanne, Chemin Pegasi 51, 1290 Versoix, Switzerland*
Email and postal address for correspondence: g.a.verdoes.kleijn@rug.nl, Kapteyn Astronomical Institute, Postbus 800, 9700 AV, Groningen, The Netherlands.

***Keywords:*** *Artificial Intelligence, NEO Discovery, NEO Characterisation, Astronomical Surveys, Polarimetry*

# 1 Introduction

We are a group of over two dozen astronomers, computer scientists, data scientists and digital Big Data research platform experts at 11 universities and research institutes in South Africa and Europe. We study Near-Earth Objects (NEOs) for Planetary Defence and scientific purposes.

We present our research and development programme for algorithms and digital data analysis platforms for machine learning-assisted NEO discovery and polarimetric characterisation in astronomical surveys. Typically, this is serendipitous because these surveys are designed for galactic and extragalactic science.

The morphological appearance of NEOs in astronomical images ranges from a point spread function of just a few pixels in extent to an elongated linear feature spanning hundreds of pixels. Here, we focus on the detection and classification of elongated linear features using solely the astronomical images as input. We aim for a pure and complete detection and classification at the lowest possible signal-to-noise ratio for both the smallest and largest linear extents. In addition, we strive to develop a pipeline for detection and classification as a generic software component that is configurable for various surveys and can be interfaced or embedded in the associated astronomical data handling systems. We aim to make the approach applicable to survey images that vary in terms of spatial resolution, image quality, and depth.

# 2 NEO Discovery

Large area astronomical imaging surveys contain appearances of many types of linear, streak-like features. These include NEOs and other solar system objects, CCD charge bleeding, diffraction spikes, cosmic ray impacts, meteor fireballs, very elongated galaxies and (increasingly) human-made satellites. Automatic detection and classification of these classes with high completeness and precision can significantly speed up the use of these surveys for solar system scientific research and Planetary Defence/Space Situational Awareness. Additionally, this detection and classification are also already valuable during the data processing and quality assessment phases.

As astronomical surveys are designed for galactic and extragalactic science, they can vary in their suitability for detecting asteroids, and NEOs in particular. Here, we focus on ongoing surveys in which our science team is involved, including various surveys conducted with OmegaCAM at the VST (e.g., Iodice 2022, Wright et al. [2024](#), [Peletier et al. 2020](#)) and the Euclid Mission surveys ([Euclid Collaboration 2024](#)). Our group also includes members of the Rubin Consortium who are interested in taking the lessons learned from Euclid and OmegaCAM to the LSST survey.

For the OmegaCAM archive, it is estimated that one in twenty NEOs appear in OmegaCAM data at a signal-to-noise ratio higher than 3 ([Saifollahi et al, 202](#)3). Of these, order 70% have elongations that make them well-resolved compared to point sources (see Table 2 and Fig.8 in Saifollahi et al 2023). For example, order 6.4E4 asteroid appearances are expected in just the Kilo-Degree Survey of OmegaCAM (Verdoes Kleijn et al 2024). Classical methods (using SourceXtractor and StreakDet) have been deployed on OmegaCAM to detect appearances of already known NEOs, i.e., precoveries (Saifollahi et al 2023). This achieves a recovery rate of 40% for NEOs on the risk list for streaks with a predicted S/N > 3, decreasing to 20% for the complete list of NEOs. The precovery rate increases to about 50% for S/N > 10. In other words, currently, the majority of NEOs with a predicted 3 < S/N < 10 remain undetected (even after visual inspection). Especially for Planetary Defence purposes, it is relevant to assess whether the failed NEO detections can indicate errors that are unaccounted for (e.g., in their photometric model).

These results make it interesting to develop a pipeline that automatically harvests NEO candidates "blindly" from the continuously increasing OmegaCAM archive and crossmatches them with the known NEO population.

The desire for a generically applicable pipeline makes it interesting to move over to machine learning assisted methods for this. Machine learning holds the promise of making it straightforward to deploy on the OmegaCAM archive, which is heterogeneous in terms of seeing properties and image depth. A machine-learning approach was done by Irureta-Goyena et al (2025). They develop a pipeline in which the Convolution Neural Network TernausNet automatically segments a single image. This network is an evolved form of a UNET architecture. They achieve a completeness of 65% for 3<S/N<20 appearances on a set of 276 visually confirmed NEO appearances in OmegaCAM r-band images with a trail length of 5-120 pixels. This appears to have a higher completeness than Saifollahi et al. (2023), but this comparison is apples-to-oranges, as the datasets are not identical. Furthermore, as the machine-learning approach is a blind search, it yields false positives, with a purity estimate of 44%.

The Euclid Wide Survey, which started nominal survey operations on 14 Feb 2024, is expected to contain approximately 15000 detectable Near-Earth Asteroids (Carry 2018). For Euclid, Pöntinen et al (2023) developed a Convolution Neural Network for asteroid detection that has six convolution layers and a dense layer. The training is done with a YOLO loss function and is based on simulated Euclid VIS images. For simulated images, it surpasses the asteroid detection completeness (recall) achieved by a traditional segmentation-based pipeline, detecting both fainter asteroids by 0.25–0.5 magnitudes and slower-moving asteroids, resulting in a 50% increase in the number of detected asteroids.

As a next step forward, we plan to develop a multi-class object detector and classifier pipeline for linear features suited for OmegaCAM, Euclid and LSST. We have started on OmegaCAM. For the asteroid streaks, this project draws inspiration from Irureta-Goyena et al. (2025) and Pöntinen et al. (2023). It shall be generalised from asteroids to detecting and classifying satellites in OmegaCAM images. We take inspiration from Stoppa et al. (2024) and Paillassa et al. (2020). The following steps after this are to generalise to more classes of linear features and to generalise to the Euclid and LSST surveys. We will evaluate the use of Vision Transformers.

# 3 NEO Polarimetric Characterisation

VSTPOL, the polarimetric mode of OmegaCAM 1 square degree imager at the VLT Survey Telescope, is planned to be commissioned in 2026 ([Schipani et al 2024](#)). For NEOs, the linear polarisation signal depends on the refractive index of the surface material (and the angle between the incident and scattered Solar light of each scattering). NEOs can be viewed under large phase angles, sometimes exceeding 90°, when polarisation levels may reach values of 10%-30% ([Bagnulo et al 2024](#)). This allows one to distinguish between a large and dark object and a small and bright object by combining VST's optical polarimetric results with optical photometry. VSTPOL is helpful in quickly constraining the size and composition of Near-Earth Objects for Planetary Defence purposes. Polarimetry allows the characterisation of an object's composition and size by combining a few polarimetric measurements with photometry. VSTPOL's combination of polarimetric accuracy and large field of view allows it to characterise new Near-Earth Objects with limited orbital knowledge. This is particularly relevant in case of significant probability of collision with Earth in the short term.

# 4 Astronomical Science with NEOs

We plan to use our NEO and general asteroid astrometry, photometry and polarimetry also for astronomical science. Precision absolute astrometry by itself can help put constraints on theories of non-standard gravity using the asteroids as "test particles" in the Solar System gravitational field ([Tsai et al. 2023](#)).

Combining optical polarimetry and photometry with thermal infrared photometry plus optical/near-IR spectroscopy can lift in detail the degeneracies between object size and albedo to provide definitive constraints on surface topology, physical composition and object size ([Bagnulo et al 2024](#)). This can help to constrain, for example, the role of NEO impacts as bringers of prebiotic material (Oba et al., 2022) and inducers of climate change (Brugger et al., [2017](#)).

# 5 NEOs and Astronomical Research Data Platforms

We want to develop our pipeline for the detection and classification of NEOs (and asteroids in general) as a generic software component that is straightforward to configure for different surveys and can be interfaced or embedded in various astronomical data handling systems. Furthermore, we aim to integrate our machine-learning-based pipelines with existing Big Data research platforms in a manner that minimises human overhead ("babysitting", manual steps), is performant (yielding quick results), and is environmentally friendly (reducing energy consumption through efficient computations and operations). One of our partners is the OmegaCEN Astronomical Science Data Centre at the University of Groningen[1].

---

[1] [https://www.rug.nl/research/kapteyn/onderzoek/areas/instrumentation_data](https://www.rug.nl/research/kapteyn/onderzoek/areas/instrumentation_data)

This centre plays a leading role in digital Big Data platforms for ESA ([Euclid Mission](#)[2]) and for ESO instrumentation ([VST-OmegaCAM](#), [VLT-MUSE](#), ELT-[MICADO](#)/[METIS](#)). These platforms build on their [AstroWISE](#) Information System, which was developed for astronomical research ([Begeman et al. 2013](#)). Therefore, at least for these systems, we are well-positioned to establish a NEO discovery and characterisation pipeline that interfaces seamlessly with these platforms and their extensive data archives, compute clusters, and databases.

From pathfinder projects involving the expertise of ESA's Planetary Defence and NEO Coordination Centre and the OmegaCEN team, we have learned the following lessons from taking this piggyback approach of connecting a NEO pipeline to a Big Data research data platform whose architecture is driven by galactic/extragalactic science. The synergies between NEO investigations and astronomical science lie in (i) standard instrument and software requirements on astrometric and photometric precision calibration and (ii) standard requirements on databasing and IT to handle such large datasets. The challenges lie in bridging the gap between communities and the sometimes non-natural fit with the traditional tasks of universities and science funding agencies. One way we address this challenge is by linking the astronomical databases to the open-source Tudat orbit estimation software for automated ephemeris updates of target Near-Earth Objects (NEOs) and dynamical validation of new observations. For this we have submitted a funding application.

We close by providing a pointer to relevant publications involving our expertise group available via the Astrophysics Data System: [NEO Planetary Defence library](#).

---

[2] see also [Euclid Netherlands Science Data Center](#)